**Trust in Generative AI for Health Information Consumption and the Effect of Learned Dependency: An Experimental Investigation**

[a]Arif Ahmed, [a] Gondy Leroy, [a]Agrim Sachdeva, [a]Philip Harber,

[a] Stephen A. Rains, [a] Seokjun Youn, [a] Prosanta Barai

[a] The University of Arizona, Tucson 85721, U.S.A

## Abstract

Background: Generative artificial intelligence (GenAI) is increasingly used by health information consumers to interpret medical content and support decision-making. While these systems provide accessible, timely information, they may also produce inaccurate or misleading outputs. Effective use of GenAI, therefore, depends on users' ability to calibrate trust based on information accuracy. However, little is known about how learned dependency (The habitual reliance on AI systems for solving problem) on GenAI influences trust calibration in health information contexts.

Objective: This study examines how learned dependency on GenAI affects health information consumers' trust calibration in AI-generated information and whether visual attention cues (e.g., highlighting critical information on text) mitigate overreliance on incorrect outputs.

Methods: We conducted two randomized controlled experiments, the first involving 338 college students and the second replicating it with 563 Amazon Mechanical Turk participants. Both studies employed a 2 × 2 between-subjects design manipulating (1) information accuracy (correct vs incorrect) and (2) visual attention cues (text highlight vs no text highlight). Participants evaluated AI-generated health information presented alongside source text. Trust was measured using a multi-item scale, and learned dependency on GenAI was assessed using a validated self-reported measure. Linear regression models were used to examine main and interaction effects.

Results: Across both experiments, information accuracy had a significant positive effect on trust, with participants expressing greater trust in correct than in incorrect AI-generated information ($p < .001$). Learned dependency was positively associated with trust in both experiments ($p < .05$), such that more dependent users trusted AI outputs more overall. Critically, the interaction between information accuracy and learned dependency was negative and significant in both experiments ($p < .001$), indicating that higher dependency reduces users' sensitivity to information accuracy. Text highlighting did not significantly affect trust in either experiment, nor did it moderate the relationship between learned dependency and trust.

Conclusions: This study demonstrates that while users generally trust accurate AI-generated health information more than inaccurate information, higher self-reported learned dependency is associated with weaker trust calibration, predicting greater susceptibility to incorrect outputs. Visual attention cues such as text highlighting alone are insufficient to mitigate this effect. These findings highlight the need for more effective design interventions to support critical evaluation and reduce overreliance on GenAI in health information environments.

# Introduction

Artificial Intelligence (AI), particularly Generative AI (GenAI), has rapidly transformed the healthcare landscape, offering new tools for patient education, clinical decision support, and personalized medical recommendations. Since the release of ChatGPT in late 2022, public engagement with GenAI systems has surged, with a growing number of individuals turning to AI-powered chatbots for health-related queries [1, 2]. These tools promise convenience and accessibility but also produce erroneous outputs, raising critical questions about how users evaluate and calibrate trust in AI-generated health information [3-13, 27]. Studies demonstrate that AI chatbot responses to health queries frequently deviate from clinical guidelines, contain inaccuracies, and may mislead patients who are unaware of these limitations [30, 31]. A key concern emerging from this pattern of use is that repeated and largely uncritical engagement with GenAI may foster overreliance. As users experience GenAI outputs as convenient and immediately accessible, they may progressively reduce their own evaluative effort, which is a process that, over time, can give rise to learned dependency. Learned dependency is a behavioral phenomenon defined as habitual reliance on external AI systems arising from repeated reinforcement, often at the expense of independent thinking and critical evaluation [14]. Research on overreliance on AI demonstrates that users frequently accept incorrect AI outputs, particularly when they perceive the system as competent [22, 25, 29, 35, 36]. This pattern, known as automation bias, is strengthened through repeated reliance, as habitual use of AI systems progressively reduces users' critical scrutiny of system outputs [22, 36]. In health information contexts, where the consequences of misplaced trust may directly affect health decisions, understanding how learned dependency shapes trust calibration is a pressing but underexplored concern.

Despite the rapid integration of GenAI into health information ecosystems, existing research has not adequately examined how learned dependency shapes trust calibration in this context. Prior work has explored dependency-related phenomena among students [15] and professionals such as clinicians or software developers [16], yet health information consumers whose GenAI use directly bears on health comprehension and decision-making has remained largely overlooked. Prior research on online health information has extensively documented how source credibility, interface design, and information quality shape user trust [34], yet the role of habitual AI dependency in disrupting trust calibration has not been experimentally examined. Critically, no prior experimental study has tested whether learned dependency disrupts users' ability to distinguish accurate from inaccurate AI-generated health information, or whether interface-level interventions such as visual attention cues can counteract such effects. Addressing this gap is essential for designing AI-mediated health systems that enhance rather than erode users' critical evaluation [17].

Our study addresses this gap through two controlled experiments examining how learned dependency on GenAI affects trust calibration in AI-generated health information. We focus on trust as the key outcome because it is the primary cognitive mechanism through which users decide whether to accept or question AI-generated content, and because prior research consistently identifies trust as central to health information engagement and AI adoption [3, 18,19, 32, 33]. We further test whether text highlighting, a practical interface-level attention cue, can mitigate dependency-driven overreliance on inaccurate AI outputs. By focusing specifically on trust as the outcome variable, this study makes a deliberate and bounded contribution: we do not examine downstream behavioral intentions or adoption behavior, which represent important but distinct constructs warranting dedicated investigation in future work.

## Theoretical Framework

Our study draws on two complementary theoretical frameworks: trust calibration theory and attention theory. Trust calibration theory holds that effective human–automation interaction requires users to adjust their trust in proportion to a system's actual reliability [18]. Well-calibrated trust enables users to rely on a system when it performs correctly and withhold trust when it errs. Critically, miscalibrated trust in either direction carries consequences: overreliance leads users to accept erroneous outputs, while underreliance causes them to discard useful information [18, 19]. This framework underpins our prediction about the connection between information accuracy and trust.

Building on these theoretical foundations, we introduce learned dependency as a behavioral phenomenon that may disrupt trust calibration. Learned dependency refers to the tendency for users to increasingly rely on AI-generated outputs after repeated exposure and perceived successful use. This construct is theoretically grounded in two related literatures. First, research on automation bias demonstrates that repeated reliance on automated systems leads users to defer to system outputs with reduced critical scrutiny, even when errors are present [22, 29, 35]. This effect is particularly pronounced in health contexts, where AI-generated advice has been shown to significantly reduce diagnostic accuracy when incorrect [35]. Second, cognitive offloading theory suggests that users who habitually delegate information evaluation to external systems progressively reduce their own cognitive investment in that task [20, 36], a pattern that becomes self-reinforcing over repeated interactions. Together, these mechanisms suggest that learned dependency may be positively associated with overall trust in AI outputs while simultaneously weakening users' sensitivity to information accuracy, as dependent users process AI outputs with reduced analytical engagement [23].

Against this backdrop, interface-level attention cues represent a practical and scalable intervention to support trust calibration. Highlighting key information increases its perceptual salience, which, according to the limited capacity model of mediated message processing [21], should free attentional resources for evaluative processing rather than information search. Eye-tracking research has shown that users consistently direct their visual attention toward highlighted text areas, demonstrating the von Restorff isolation effect for digital highlighting interfaces [37, 39]. The Von Restorff effect, also known as the isolation effect, suggests that when people are presented with several similar items, the item that stands out as different is more likely to capture attention and be remembered. In digital highlighting interfaces, this principle works by making highlighted information visually distinct, prompting users to pause and process it more carefully. Importantly, among available highlighting methods, bold text has been found to produce the shortest visual search times in graphical user interface tasks, outperforming color-only highlighting on search efficiency [40]. Whether such attentional cueing is sufficient to support trust calibration among highly dependent users, however, remains an open empirical question. We posit our research questions as:

RQ1: How do the accuracy of AI-generated health information and visual attention cues (e.g., highlighting text information) influence health information consumers' trust in AI-generated information?

RQ2: How does learned dependency on generative artificial intelligence influence health information consumers' trust in AI-generated information, and how does it moderate the effects of information accuracy and visual attention cues on trust?

## Hypotheses Development and Research Framework

Building on the integrative framework presented above, we examine how information accuracy, visual attention cues, and learned dependency on GenAI jointly shape health information consumers' trust in AI-generated content. Rooted in trust calibration theory and attention theory, our framework proposes that habitual reliance on GenAI reduces scrutiny in information evaluation, with important consequences for how users respond to accurate and inaccurate AI-generated outputs. We have developed the research framework (See Figure 1) and corresponding hypotheses presented below.

Trust calibration theory predicts that users should, under normal conditions, trust accurate information more than inaccurate information. When AI generated health information is correct, users who evaluate it critically are expected to recognize its quality and report higher trust. We therefore hypothesize:

H1: Information accuracy positively associate health information consumers' trust in AI generated health information.

Attention theory suggests that directing users' attention to specific portions of AI generated health information through visual cues can support more careful evaluation of that content. In this study, text highlighting is applied to key portions of the AI generated health text regardless of whether those portions contain accurate or inaccurate information, depending on the experimental condition. By directing attentional resources toward the highlighted content, users are better positioned to compare it against the source clinical text and detect potential inaccuracies. Text highlighting is therefore proposed as a design intervention that supports trust calibration rather than simply increasing or decreasing trust such as users who receive highlighting should be more sensitive to information accuracy than those who do not. We therefore hypothesize:

H2: The presence of visual attention cues (text highlighting) is positively associated with health information consumers' trust in AI generated health information.

Research on automation bias and cognitive offloading suggests that users who have developed habitual reliance on AI systems tend to engage with AI outputs with reduced analytical effort. As a result, highly dependent users are expected to extend greater overall trust to AI generated information, regardless of its accuracy. We therefore hypothesize:

H3: Learned dependency on GenAI is positively associated with health information consumers' trust in AI generated health information.

While H3 predicts a positive main effect of learned dependency on trust, trust calibration theory further predicts that this dependency may distort the relationship between information accuracy and trust. Specifically, users with high learned dependency are expected to process AI outputs with reduced scrutiny, weakening their ability to distinguish accurate from inaccurate content. We therefore hypothesize:

H4: Higher learned dependency is negatively associated with a positive relationship between information accuracy and health information consumers' trust in AI generated health information, such that the effect of information accuracy on trust is weaker when learned dependency is higher.

Finally, while visual attention cues may generally support trust calibration (H2), their effectiveness may be attenuated among users with high learned dependency. Highly dependent users may continue to accept AI outputs with limited scrutiny even when key information is visually highlighted, as habitual reliance reduces their motivation to engage in effortful evaluation. We therefore hypothesize:

H5: Visual attention cues weaken the trust-distorting effect of learned dependency on trust calibration, such that the negative moderating effect of learned dependency on the relationship between information accuracy and trust (H4) is attenuated when visual attention cues are present.

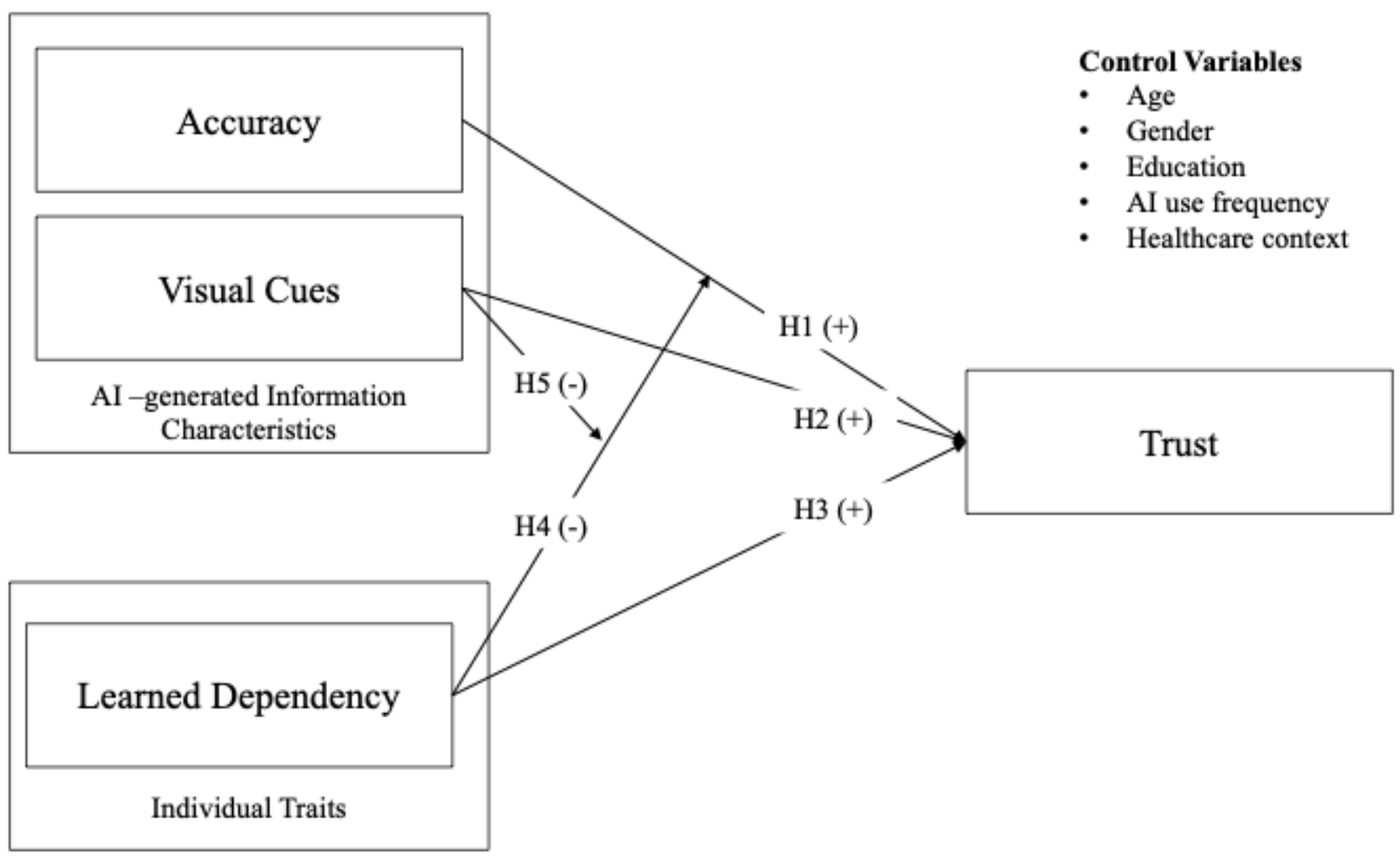

Figure 1: Research Framework

Figure 1 illustrates the research framework guiding this study. Information accuracy and visual attention cues are positioned as AI generated information characteristics that directly influence trust. Learned dependency is positioned as an individual trait that both directly influences trust (H3) and moderates the effects of information accuracy (H4) and visual attention cues (H5) on trust. The framework reflects a 2x2 experimental design in which information accuracy and visual attention cues are manipulated between subjects, while learned dependency is measured as a continuous individual difference variable.

Our study contributes to the literature in several ways. First, it advances research on human–AI interaction by introducing learned dependency as a behavioral phenomenon that can influence trust calibration in AI-generated health information. While prior work has examined trust in AI systems [24-26], automation bias [22, 29], cognitive forcing functions [25], and GenAI dependency [17] separately, no prior study has experimentally tested their joint influence on trust calibration in health information contexts.

Second, the study tests whether text highlighting is a practical interface-level attention cue that can attenuate the trust-distorting effect of learned dependency. Building on prior work on cognitive forcing functions [25] and visual attention cues [39, 40], the study extends this literature to GenAI health information contexts and provides evidence on the limits of surface-level attentional interventions.

Third, the study offers practical implications for the design of generative AI systems used in health information environments. As consumers increasingly rely on GenAI tools for medical guidance and interpretation of health information, understanding how trust and dependency shape user behavior becomes critical for the safety of health information consumers [3-6]. The findings highlight the need for interface designs and decision-support mechanisms that promote critical evaluation and reduce the risk of overreliance on AI-generated health information.

Collectively, these contributions deepen our understanding of how GenAI shapes trust calibration in health information contexts, positioning learned dependency as a key mechanism through which habitual AI use influences users' evaluative judgment.

# Methodology

We conducted two randomized controlled experiment to examine the impact of GenAI on human trust calibration. The experiment evaluated participants' trust in AI-generated health information using a between-subjects design with randomized assignment to conditions.

## Ethical Considerations

Our study was approved by the Institutional Review Board (IRB) prior to data collection (IRB Protocol: STUDY00002235). All participants provided informed consent before beginning the survey, covering the study's purpose, estimated completion time, voluntary participation, and data storage procedures. No personally identifiable information was collected. Data were stored securely on the Qualtrics platform and accessible only to the research team. The health information stimuli were sourced from the MIMIC-III clinical database, a credentialed-access dataset, under a data use agreement held by the research team.

## Experiment 1

### Recruitment of Participants

Experiment 1 employed a convenience sampling approach. Participants were junior-level students enrolled in a business course at a university. The use of junior-level business students is appropriate because trust calibration and learned dependency are not domain-specific constructs [14, 17]. The course instructor informed students about the study in advance, and participants received 10 extra credit points upon completion. The study was conducted online, beginning with a pre-task demographic questionnaire that collected data on participants' age, gender, education level, race, ethnicity, and major field of study. In addition, participants completed a six-item GenAI dependency scale, adapted from prior research on media and technology reliance [28] (See Multimedia Appendix A for the full item list). Responses were recorded on a five-point Likert scale ranging from strongly disagree (1) to strongly agree (5), with the items designed to assess participants' habitual reliance on GenAI tools for decision-making and information gathering. Following the initial survey, participants were assigned to the experimental condition, which assessed trust in AI-curated health information. All instructions and materials were delivered electronically through the study platform (Qualtrics[1]).

## Experiment Design

This experiment employed a 2×2 between-subjects factorial experimental design to examine how information accuracy and visual attention cues influence health information consumers' trust calibration in AI-generated health information, and how learned dependency on GenAI moderates these relationships. The two independent variables were: (1) information accuracy (correct vs. incorrect AI-generated health information) and (2) visual attention cues (text highlighting via bold text vs. no highlighting). The dependent variable was trust in AI-generated health information. Learned dependency was measured as a continuous individual difference variable. Participants were randomly assigned to one of 20 experimental groups (2 accuracy conditions × 2 visual cue conditions × 5 distinct health texts), with randomization handled automatically by Qualtrics. Using five texts reduces the risk that observed effects are an artifact of one particular stimulus, increasing confidence that the accuracy and highlighting effects generalize across content.

[1] https://uarizona.co1.qualtrics.com/

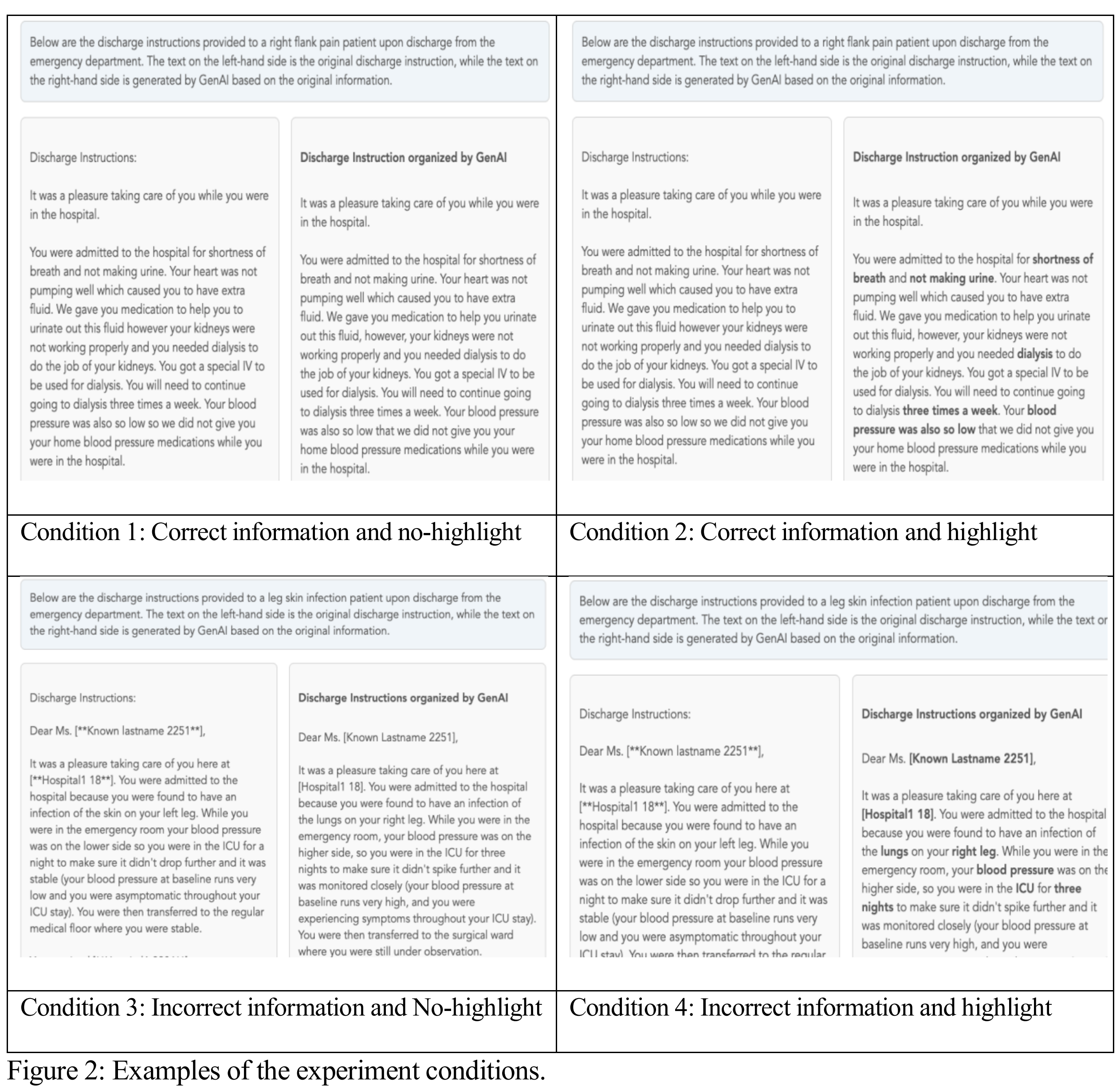
Below are the discharge instructions provided to a right flank pain patient upon discharge from the emergency department. The text on the left-hand side is the original discharge instruction, while the text on the right-hand side is generated by GenAI based on the original information.
Discharge Instructions:
It was a pleasure taking care of you while you were in the hospital.
You were admitted to the hospital for shortness of breath and not making urine. Your heart was not pumping well which caused you to have extra fluid. We gave you medication to help you to urinate out this fluid however your kidneys were not working properly and you needed dialysis to do the job of your kidneys. You got a special IV to be used for dialysis. You will need to continue going to dialysis three times a week. Your blood pressure was also so low so we did not give you your home blood pressure medications while you were in the hospital.
Discharge Instruction organized by GenAI
It was a pleasure taking care of you while you were in the hospital.
You were admitted to the hospital for shortness of breath and not making urine. Your heart was not pumping well which caused you to have extra fluid. We gave you medication to help you urinate out this fluid, however, your kidneys were not working properly and you needed dialysis to do the job of your kidneys. You got a special IV to be used for dialysis. You will need to continue going to dialysis three times a week. Your blood pressure was also so low that we did not give you your home blood pressure medications while you were in the hospital.
Condition 1: Correct information and no-highlight
Condition 2: Correct information and highlight
Below are the discharge instructions provided to a leg skin infection patient upon discharge from the emergency department. The text on the left-hand side is the original discharge instruction, while the text on the right-hand side is generated by GenAI based on the original information.
Discharge Instructions:
Dear Ms. [**Known lastname 2251**],
It was a pleasure taking care of you here at [**Hospital1 18**]. You were admitted to the hospital because you were found to have an infection of the skin on your left leg. While you were in the emergency room your blood pressure was on the lower side so you were in the ICU for a night to make sure it didn't drop further and it was stable (your blood pressure at baseline runs very low and you were asymptomatic throughout your ICU stay). You were then transferred to the regular medical floor where you were stable.
Discharge Instructions organized by GenAI
Dear Ms. [Known Lastname 2251],
It was a pleasure taking care of you here at [Hospital1 18]. You were admitted to the hospital because you were found to have an infection of the lungs on your right leg. While you were in the emergency room, your blood pressure was on the higher side, so you were in the ICU for three nights to make sure it didn't spike further and it was monitored closely (your blood pressure at baseline runs very high, and you were experiencing symptoms throughout your ICU stay). You were then transferred to the surgical ward where you were still under observation.
Condition 3: Incorrect information and No-highlight
Condition 4: Incorrect information and highlight
Figure 2: Examples of the experiment conditions.

Stimuli

Discharge instructions were sourced from the MIMIC-III[2] clinical database, a credentialed-access de-identified dataset widely used in clinical research. Five summaries were randomly selected to represent common inpatient medical conditions. Each original summary was presented alongside a version generated by GPT-4o[3] (default version, default settings). For the accuracy manipulation, GPT-4o-generated summaries either preserved the original clinical content (correct condition) or contained deliberate alterations in critical medical details (incorrect condition): three locations in the text. In the visual attention cue condition, key portions of the AI-generated text were highlighted using bold text [39, 40]. No formal manipulation check was included, as doing so would contaminate the trust measurement. This is acknowledged as a limitation.

Procedure

Participants completed the study in approximately 10 minutes, with a maximum time limit of 20 minutes. In the trust evaluation task, participants were presented with two texts side by side: the original MIMIC-III discharge instruction and the corresponding GPT-4o-generated version (See Figure 2). The AI-generated version varied in two key ways depending on the assigned condition: (1) accurate or inaccurate information and (2) presence or absence of bold-text highlighting. Qualtrics automatically randomized participants to one of the twenty experimental groups, with back-navigation disabled. After reviewing both texts, participants completed the trust scales.

[2] https://physionet.org/content/mimiciii/1.4/
[3] https://openai.com/index/gpt-4-research/

### Measures

Trust in AI-generated health information was measured using a three-item scale adapted from Lee and See [18] and modified for the GenAI health context: (1) "The GenAI generated information is trustworthy," (2) "The GenAI's output is accurate," and (3) "The GenAI is well prepared to handle public health data." Responses were on a five-point Likert scale (1 = Strongly Disagree to 5 = Strongly Agree). The scale demonstrated acceptable reliability (α = .724, CR = .729). Learned dependency was mean-centered prior to constructing interaction terms.

## Experiment 2

### Overview

To enhance the external validity of Experiment 1 and extend the findings to a more diverse population, we conducted a replication study (Experiment 2) using Amazon Mechanical Turk (MTurk). Experiment 2 retained the same 2×2 between-subjects factorial design (information accuracy × visual attention cues) and tested an identical set of hypotheses (H1–H5). Two key differences were introduced to address limitations of Experiment 1. First, the learned dependency scale was expanded from 6 to 17 items to improve psychometric coverage of the construct. Second, the accuracy manipulation was strengthened from three to six altered locations within the AI-generated text. Third, participants were recruited from MTurk rather than a university setting, enabling a broader and more occupationally diverse sample.

Recruitment of Participants

Participants were recruited via Amazon Mechanical Turk (MTurk), a widely used crowdsourcing platform for behavioral research. Eligible participants were required to be US-based MTurk workers with an approval rating above 95%, and were compensated $0.50 upon completion. All instructions and materials were delivered electronically through the Qualtrics.

Experiment Design

Experiment 2 retained the same 2×2 between-subjects factorial design as Experiment 1, manipulating information accuracy (correct vs. incorrect) and visual attention cues (text highlighting vs. no highlighting), with participants randomly assigned to one of four conditions across five health texts (20 experimental groups). The accuracy manipulation was strengthened relative to Experiment 1: incorrect information was introduced in six locations within the AI-generated health text (compared to three in Experiment 1), with all six locations highlighted in the visual cue condition.

Stimuli

The same MIMIC-III-derived discharge instructions and GPT-4o-generated counterparts used in Experiment 1 served as stimuli, with the accuracy manipulation extended to six altered locations per text (compared to three in Experiment 1) to create a stronger experimental treatment.

Procedure

The procedure mirrored Experiment 1: participants viewed the original discharge instructions alongside the GPT-4o-generated version, varying in accuracy and presence/absence of bold-text highlighting, with condition assignment automatically randomized by Qualtrics.

Measures

Trust in AI-generated health information was measured using a four-item scale (α = .796, CR = .800, AVE = .501), following confirmatory factor analysis (CFA) which identified and removed two items with weak standardized loadings from the original six-item pool, which combined three items adapted from Lee and See [18] with three items adapted from Turel [19]. The retained items were "The GenAI's output is accurate," "The GenAI is well prepared to handle public health data," "I trust AI tools," and "AI tools are trustworthy." The learned dependency on GenAI construct was assessed using an expanded 17-item scale combining the six items from Experiment 1 with 11 items adapted from Goh et al. [17], capturing a broader range of dependency-related behaviors and attitudes (α = .932, CR = .930). Responses for both scales were recorded on a five-point Likert scale (1 = Strongly Disagree to 5 = Strongly Agree). Learned dependency was mean centered prior to constructing interaction terms.

## Reliability & Validity

### Trust Measure Items

In Experiment 1, the three-item trust scale adapted from Lee and See [18] demonstrated acceptable internal consistency (Cronbach's α = .724, Composite reliability (CR) = .729). Because a three-item single-factor model is just-identified (df = 0), global fit indices cannot be meaningfully assessed; standardized factor loadings ranged from 0.594 to 0.808 (AVE = 0.478). The AVE falls marginally below the 0.50 threshold, which is acceptable when CR exceeds 0.70 (Fornell and Larcker, 1981). In Experiment 2, the trust scale was expanded to six items by adding three items adapted from Turel [19]. Following CFA, items 1 and 4 were removed due to weak standardized loadings (0.511 and 0.420, respectively). The retained four-item scale demonstrated excellent model fit ($\chi^2$/df = 0.806, CFI = 1.000, RMSEA = .000) and strong psychometric properties ($\alpha$ = .796, CR = .800, AVE = .501). Full CFA results for both experiments are reported in Multimedia Appendix D.

Learned Dependency on GenAI Items

In Experiment 1, the six-item learned dependency scale adapted from prior research on technology reliance [28] demonstrated good model fit ($\chi^2$/df = 2.777, CFI = .977, RMSEA = .068) and acceptable reliability ($\alpha$ = .782, CR = .786, AVE = 0.416), though items 1 and 2 showed weak standardized loadings (0.109 and 0.489 respectively), noted as a limitation. In Experiment 2, the scale was expanded to 17 items by combining the six Experiment 1 items with 11 items adapted verbatim from Goh et al. [17]. This expanded scale demonstrated excellent internal consistency and composite reliability ($\alpha$ = .932, CR = .930) with acceptable model fit (CFI = .915, RMSEA = .082), consistent with the original validation by Goh et al. [17]. The AVE (0.462) falls marginally below the 0.50 threshold, which is common for large item pools where individual items capture distinct behavioral facets of the construct. Response distributions showed minimal straight-lining (0.8% for the dependency scale; 1.0% for the trust scale), and all variance inflation factors were below 3.2, indicating no multicollinearity concerns. Discriminant validity was assessed using the Fornell-Larcker criterion by comparing the inter-construct correlation between Trust and Learned Dependency against the square root of AVE for each construct (see Table D4). In Experiment 1, discriminant validity was not established (inter-construct r = .697 exceeded √AVE for both constructs), indicating overlap between the trust and dependency measures in this sample. In Experiment 2, discriminant validity was supported (r = .643, below √AVE for both constructs), following CFA-based refinement of both scales. Full CFA results are reported in Multimedia Appendix D.

# Results

## Demographics

A total of 338 participants completed the 1st experimental study (see Table 1). The sample was nearly evenly split by sex (50.59% male, 49.41% female) and predominantly aged 18–30 (99.41%). Most participants identified as White (78.70%), and 20.41% identified as Hispanic or Latino. Educational attainment was relatively low, with 80.47% having completed only high school. Reflecting the recruitment context, 97.93% of participants were Business or Economics majors.

A total of 563 participants completed the 2nd experimental study (see Table 1). The sample was predominantly male (77.98%) and primarily aged 21–40 (94.14%), with a mean age of 31.7 years. Most participants identified as White (95.56%), and 23.62% identified as Hispanic or Latino. Educational attainment was considerably higher than in Experiment 1, with 72.82% holding a bachelor's degree and 22.20% a master's degree. Major fields of study were more varied, led by STEM (31.97%), Health or Medicine (19.54%), and Business or Economics (19.36%).

Table 1: Demographic information.

| | Experiment 1 N=338 (%) | Experiment 2 N=563 (%) |
|---|---|---|
| **Gender** | | |
| Male | 171 (50.59) | 439 (77.98) |
| Female | 167 (49.41) | 122 (21.67) |
| Other | 0 (0) | 2 (0.36) |
| **Age** | | |
| 18 to 20 years old | 102 (30.18) | 1 (0.18) |
| 21 to 30 years old | 234 (69.23) | 348 (61.81) |
| 31 to 40 years old | 2 (0.59) | 182 (32.33) |
| 41 to 50 years old | 0 (0) | 13 (2.31) |
| 51 years and above | 0 (0) | 19 (3.37) |
| **Race** | | |
| White | 266 (78.70) | 538 (95.56) |
| Asian | 29 (8.58) | 7 (1.24) |
| Black | 7 (2.07) | 7 (1.24) |
| American Indian | 3 (0.89) | 3 (0.53) |
| Hawaiian | 0 (0) | 1 (0.18) |

| More than one race | 23 (6.80) | 0 (0) |
|---|---|---|
| American Indian, White | 0 (0) | 4 (0.71) |
| White, More than one race | 0 (0) | 2 (0.36) |
| Asian, White | 0 (0) | 1 (0.18) |
| Black, White | 0 (0) | 1 (0.18) |
| Unknown | 3 (0.89) | 0 (0) |
| **Ethnicity** | | |
| Hispanic or Latino | 69 (20.41) | 133 (23.62) |
| Not Hispanic or Latino | 263 (77.81) | 426 (75.67) |
| Unknown | 6 (1.78) | 4 (0.71) |
| **Education** | | |
| Less Than High School | 0 (0) | 2 (0.36) |
| High School | 272 (80.47) | 3 (0.53) |
| Associate's Degree | 43 (12.72) | 13 (2.31) |
| Bachelor's Degree | 23 (6.80) | 410 (72.82) |
| Master's Degree | 0 (0) | 125 (22.20) |
| Other Professional Degree | 0 (0) | 1 (0.18) |
| Doctorate | 0 (0) | 9 (1.60) |
| **Major Field of Study** | | |
| Science, Technology, Engineering, and Mathematics | 4 (1.18) | 180 (31.97) |
| Health or Medicine | 0 (0) | 110 (19.54) |
| Business or Economics | 331 (97.93) | 109 (19.36) |
| Social Sciences or Humanities | 0 (0) | 66 (11.72) |
| Arts or Design | 0 (0) | 39 (6.93) |
| Education or Teaching | 0 (0) | 48 (8.53) |
| Law or Criminology | 0 (0) | 5 (0.89) |
| Agriculture or Environmental Studies | 0 (0) | 2 (0.36) |
| Other | 3 (0.89) | 4 (0.71) |

## Data Analysis

All statistical analyses were conducted using R version 4.2.1. Hierarchical multiple regression (HMR) was the primary analytical approach, with variables entered in three blocks: Block 1 included control variables (healthcare context, age, education, and gender); Block 2 added the main effects (information accuracy, visual attention cues, and learned dependency); and Block 3 added the two interaction terms (Information Accuracy × Learned Dependency; Visual Attention Cue × Learned Dependency). Learned dependency was mean-centered prior to constructing interaction terms to reduce multicollinearity. Incremental model fit was assessed using $\Delta R^2$ and the associated F-test at each block. Common method bias was assessed using Harman's single-factor test. To assess the psychometric properties of the multi-item constructs, a confirmatory factor analysis (CFA) was estimated for the trust scale and the learned dependency scale using the lavaan package in R. Model fit was evaluated using standard indices including SRMR, CFI, TLI, and RMSEA, and convergent and discriminant validity were assessed using standardized factor loadings, AVE, composite reliability, and Cronbach's alpha. Full CFA results including factor loadings and fit indices are reported in the Multimedia Appendix."

## Regression

To examine the effects of information accuracy, visual attention cues, and learned dependency on trust in AI-generated health information, we conducted linear regression analyses. Table 2 presents the regression coefficients, and Table 3 summarizes hypothesis testing results. Experiment 1 (See Table 2) result consistent with H1, information accuracy had a significant positive effect on trust (B = 2.107, 95% CI [1.337, 2.878], $p < .001$). The results show that participants reported substantially higher trust in accurate AI-generated information than in inaccurate AI-generated information, indicating that, on average, results are consistent with users being more sensitive to information accuracy between correct and incorrect outputs. In contrast, visual attention cues (highlighting) did not have a significant effect on trust (B = 0.149, 95% CI [-0.622, 0.920], $p = .704$), providing no support for H2. This suggests that directing users' attention through highlighting does not meaningfully influence their trust in AI-generated information. Supporting H3, learned dependency on GenAI was positively associated with trust (B = 0.277, 95% CI [0.033, 0.521], $p = .026$). Individuals with higher levels of dependency exhibited greater overall trust in AI-generated outputs. Among the control variables, most effects were not statistically significant. However, exposure to certain text conditions (Healthcare Context4: B = - 0.449, p = .012; Healthcare Context 5: B = -0.610, p < .001) was associated with lower trust, suggesting potential contextual-related effects.

To assess whether learned dependency influences trust calibration, interaction terms were included in the model (See Table 2). Consistent with H4, the interaction between information accuracy and learned dependency was negative and significant (B = -0.399, 95% CI [-0.695, -0.104], p = .008). This finding indicates that higher dependency exhibit reduced sensitivity to information accuracy. In other words, highly dependent users are more likely to trust incorrect AI-generated information than less dependent users, reflecting a pattern consistent with diminished trust calibration, as described in the automation bias literature [29]. In contrast, the interaction between visual attention cues, learned dependency, and accuracy was not significant (B = -0.023, 95% CI -0.618, 0.573], p= .941), providing no support for H5 (See Table 3). This suggests that attention-based interventions, such as highlighting critical information, do not mitigate the effect of dependency on trust. Overall, the results provide strong support for the role of learned dependency in shaping trust in AI-generated health information.

Table 2: Regression Results.

| | Experiment 1 | | | Experiment 2 | | |
|---|---|---|---|---|---|---|
| **Variable** | **B** | **95% CI** | **P** | **B** | **95% CI** | **P** |
| Intercept) | 2.213 | 0.615 to 3.812 | 0.007 | 3.624 | 3.337 to 3.911 | < .001 |
| **Control** | | | | | | |
| Healthcare Context 2 | -0.230 | -0.576 to 0.116 | 0.193 | 0.003 | -0.136 to 0.142 | 0.968 |
| Healthcare Context 3 | -0.165 | -0.508 to 0.179 | 0.346 | 0.011 | -0.127 to 0.149 | 0.876 |
| Healthcare Context 4 | -0.449 | -0.797 to -0.101 | 0.012 | 0.007 | -0.145 to 0.130 | 0.918 |
| Healthcare Context 5 | -0.610 | -0.954 to -0.265 | < .001 | -0.068 | -0.208 to 0.071 | 0.335 |
| Gen AI Dependence measure | 0.277 | 0.033 to 0.521 | 0.026 | 0.822 | 0.715 to 0.929 | < .001 |
| Education: High School | 0.087 | -0.247 to 0.422 | 0.608 | 0.889 | 0.281 to 1.497 | 0.004 |
| Education: Less Than High School | 0.342 | -1.673 to 2.357 | 0.739 | 0.678 | -0.127 to 1.483 | 0.099 |
| Education: Associate's Degree | NA | NA | NA | - 0.228 | -0.520 to 0.065 | 0.127 |
| Education: Master's Degree | NA | NA | NA | 0.018 | -0.093 to 0.129 | 0.747 |
| Education: Other Professional Degree | NA | NA | NA | 0.088 | -0.948 to 1.125 | 0.867 |
| Education: Doctorate | NA | NA | NA | 0.099 | -0.449 to 0.251 | 0.578 |
| Age | 0.002 | -0.058 to 0.062 | 0.950 | 0.014 | 0.006 to 0.022 | < .001 |
| Gender | 0.005 | -0.035 to 0.082 | 0.851 | 0.017 | -0.127 to 0.093 | 0.757 |
| **Predictors** | | | | | | |
| Information: Correct | 2.107 | 1.337 to 2.878 | < .001 | 0.203 | 0.115 to 0.290 | < .001 |
| Visual Cue: No-highlight | 0.149 | -0.622 to 0.920 | 0.704 | 0.013 | -0.074 to 0.101 | 0.761 |
| **Interactions** | | | | | | |
| Information: Correct x GenAI Dependence | -0.399 | -0.695 to -0.104 | 0.008 | -0.459 | -0.577 to -0.340 | < .001 |
| Visual Cue: No-highlight x GenAI Dependence | -0.009 | -0.305 to 0.287 | 0.950 | -0.063 | -0.179 to 0.054 | 0.292 |
| Information: Correct × Visual Cue: No-highlight × GenAI Dependence | -0.023 | -0.618 to 0.573 | 0.941 | 0.142 | -0.095 to 0.379 | 0.240 |

Experiment 2 (See Table 2) replicated this pattern using the refined four-item trust scale. Consistent with H1, information accuracy again had a significant positive effect on trust (B = 0.203, 95% CI [0.115, 0.290], $p < .001$). Visual attention cues again showed no significant effect on trust (B = 0.013, 95% CI [-0.074, 0.101], $p = .761$), providing no support for H2. Supporting H3, learned dependency was positively and more strongly associated with trust than in Experiment 1 (B = 0.822, 95% CI [0.715, 0.929], $p < .001$). The interaction between information accuracy and learned dependency was again negative and significant (B = -0.459, 95% CI [-0.577, -0.340], $p < .001$), providing further support for H4 and indicating that the trust-distorting effect of learned dependency replicated in a larger, more diverse sample with a strengthened accuracy manipulation. The interaction between visual attention cues, learned dependency, and accuracy remained non-significant (B = 0.142, 95% CI -0.095, 0.379], $p = 0.240$), again providing no support for H5. Taken together, the cross-sample replication of H1, H3, and H4, alongside the consistent null findings for H2 and H5, strengthens confidence in the robustness of these effects (See Table 3).

While users generally trust accurate information more than inaccurate information, increased dependency weakens this distinction. Furthermore, text highlighting do not significantly influence trust nor reduce overreliance among highly dependent users.

Table 3: Hypotheses Results

| Hypothesis | Predicted Relationship | Direction | Experiment 1 (B) | Result | Experiment 2 B | Result |
|---|---|---|---|---|---|---|
| H1 | Information accuracy → Trust | Positive | 2.107 | Supported** | 0.203 | Supported *** |
| H2 | Visual attention cues → Trust | Positive | 0.149 | Not Supported | 0.013 | Not Supported |
| H3 | Learned dependency → Trust | Positive | 0.277 | Supported * | 0.822 | Supported *** |
| H4 | Information accuracy × Learned dependency → Trust | Negative | -0.399 | Supported ** | -0.459 | Supported *** |
| H5 | Information accuracy × Learned dependency × Visual attention cues → Trust | Positive | -0.023 | Not Supported | 0.142 | Not Supported |

Note. B = Unstandardized regression coefficient. Significance levels: *** $p < .001$, ** $p < .01$, * $p < .05$.

# Discussion

## Principal Findings

This study examined how characteristics of AI-generated health information and users' behavioral reliance on generative artificial intelligence influence trust in AI-generated health information. By integrating trust calibration theory and attention theory, our findings provide insights into how users evaluate AI-generated health information and how behavioral dependency may shape these evaluations.

First, the results suggest that information accuracy remains an important determinant of trust in AI-generated health information which is consistent with trust calibration theory [18]. Across both studies, participants generally expressed greater trust in accurate AI-generated responses than in inaccurate ones. This finding indicates that users can calibrate their trust based on the quality of the information provided by AI systems.

However, the findings also reveal that learned dependency on generative AI plays a critical role in shaping users' trust judgments. Individuals who reported higher levels of dependency on AI systems exhibited greater trust in AI-generated health information overall. More importantly, dependency weakened users' ability to differentiate between accurate and inaccurate information. This pattern suggests that dependence on AI tools may reduce users' ability to critically evaluate AI outputs, increasing the likelihood of overreliance on GenAI-generated information. From the perspective of human–automation interaction, this pattern is consistent with automation bias in which users defer to automated systems even when the outputs may be incorrect which is observed across both studies. These findings are consistent with theoretical mechanisms described in the automation bias literature [22, 29, 35] and cognitive offloading perspective [20, 36].

The results also provide insight about the effectiveness of visual attention cues in supporting users' evaluation of AI-generated health information. Although attention theory suggests that directing users' attention to critical information may improve information processing, the visual highlighting intervention did not significantly affect trust calibration across both studies. This finding suggests that interface cue such as text highlighting alone to specific portions of AI-generated information may not be sufficient to change how users evaluate its credibility including in Experiment 2 where the visual treatment was substantially strengthened (doubled). Users who already depend heavily on AI systems may continue to rely on AI output regardless of such interface cues.

Taken together, these findings highlight an important behavioral tension in AI-assisted health information environments. While users appear capable of recognizing differences in information accuracy, behavioral reliance on generative AI can weaken this evaluative process. As a result, dependency on AI tools may increase users' vulnerability to trusting inaccurate AI-generated health information. These dynamics raise important considerations for the design and governance of generative AI systems in highly sensitive contexts such as healthcare, where incorrect information may have adverse consequences for health-related decision making.

## Limitations

Several limitations should be considered when interpreting the findings of this study. First, while the inclusion of Experiment 2 using a diverse MTurk population sample substantially improves upon the student-only design of Experiment 1, MTurk samples themselves may differ from the broader general population in terms of digital literacy and online task experience. Future research should examine these relationships in clinical populations and among older adults with active health concerns. Second, the attention-based intervention across both experiments focused on text highlighting to direct users' attention to key information. Importantly, the manipulation was strengthened in Experiment 2 by highlighting six locations rather than three, yet the null finding for H5 persisted, providing strong evidence that highlighting alone is insufficient to counteract dependency-driven trust distortion regardless of intervention intensity. Future studies should test more substantive cognitive interventions such as uncertainty indicators, confidence cue, or AI reasoning explanations. Third, the scenarios used in both experiments represent controlled and simplified representations of health information. Real-world health decision-making contexts often involve higher stakes, emotional factors such as fear, urgency, stress, or hope, and richer multimodal information formats including colors, images, and video, all of which may influence how individuals interpret and trust AI-generated information. Fourth, this study focuses on trust as the proximal outcome and does not measure downstream behavioral intentions or actual health decisions, which represent important next steps for this research. Fifth, no formal manipulation check was included for the accuracy manipulation which a deliberate design choice to avoid contaminating the trust measurement but this limits independent verification of the manipulation's effectiveness. Sixth, learned dependency was measured through self-report, which may be subject to social desirability bias. Future research should complement self-report measures with behavioral or longitudinal indicators. Seventh, both experiments used a controlled side-by-side presentation that differs from naturalistic health information seeking, limiting ecological validity. Eighth, the visual attention manipulation used bold text only; the null H5 finding should be interpreted as specific to this implementation rather than as evidence that all visual attention interventions would fail. Ninth, in Experiment 1, the trust and learned dependency scales did not demonstrate discriminant validity (inter-construct correlation exceeded $\sqrt{AVE}$ for both constructs), suggesting these constructs may not have been empirically distinguishable with the shorter scales used in that experiment. This concern was resolved in Experiment 2 following CFA-based scale refinement, but it remains a limitation of the Experiment 1 measurement model and should be considered when interpreting the H3 and H4 findings from that experiment.

## Future Directions

Future research should pursue the following specific directions. First, future studies should test more substantive cognitive interventions beyond text highlighting specifically uncertainty indicators that display the AI's confidence or uncertainty alongside its output to determine whether these engage users at a deeper evaluative level. Second, longitudinal designs tracking the same users over extended GenAI use would allow examination of how dependency develops over time. Third, future studies should examine these relationships in clinical populations with active health concerns. Fourth, behavioral outcome measures such as verification-seeking should complement self-reported trust. Fifth, multimodal stimulus designs incorporating color, images, and variable typography should be tested to determine whether richer visual interventions overcome the null H5 finding. Sixth, future studies should examine how health literacy, medical knowledge, and digital literacy moderate the dependency-trust relationship, and include more diverse and clinically relevant populations.

## Conclusions

Across two controlled experiments with distinct populations, our findings provide robust empirical evidence that both characteristics of AI-generated information and users' behavioral traits influence trust calibration in generative AI systems used for health information consumption. Information accuracy positively influences trust, and learned dependency consistently increases overall trust while reducing sensitivity to information accuracy. Critically, text highlighting failed to counteract dependency-driven trust distortion in either experiment, including Experiment 2 where the visual intervention was substantially strengthened by highlighting incorrect information in six locations rather than three. This pattern of null findings, replicated under conditions of stronger treatment, constitutes a substantive finding in its own right: the trust-distorting effect of learned dependency is robust and cannot be mitigated by surface-level attentional cueing alone. These findings highlight the importance of designing AI-enabled health information systems that promote appropriate trust calibration through more substantive cognitive interventions. As generative AI tools become increasingly integrated into digital health ecosystems, understanding how dependency shapes evaluative judgment will be critical for ensuring safe and responsible use of AI-generated health information [31, 38].

## Funding Statement

The research reported in this paper was supported by the National Library of Medicine of the National Institutes of Health under Award Number R01LM011975. The content is solely the responsibility of the authors and does not necessarily represent the official views of the National Institutes of Health.

## Acknowledgements

We would also like to acknowledge the contributions of the participants of this study.

## Authors Contributions

Arif Ahmed: Writing – original draft & editing, Validation, Methodology, Conceptualization, Data curation, formal analysis. Gondy Leroy: Writing – review & editing, Supervision, Funding acquisition, Validation, Methodology. Agrim Sachdeva: Writing – review & editing, Research Framework. Philip Harber: Writing – review, Validation, Methodology, Conceptualization. Stephen A. Rains: Writing – Methodology, review, Conceptualization. Seokjun Youn: Writing – review & editing. Prosanta Barai: Writing – review & editing.

## Disclosure statement

None declared.

## Data Availability Statement

The study was conducted under University of Arizona IRB Protocol STUDY00002235, which was determined to be Exempt. The dataset doesn't contain identifiable information collected from human participants. However, according to the data availability statement that we acquired from IRB for this study prohibit us to make the data available to the general public. We have added the consent file as a supplementary material. We can only made available the study materials and survey materials and those are provided as a supplementary material.

## GenAI Usage Statement

During the preparation of this manuscript, Claude Sonnet 4.6[4] and Grammarly[5] were used solely to assist with paraphrasing and grammar checking. The authors reviewed and verified all generated content and take full responsibility for the final manuscript.

# REFERENCES

1. Esmaeilzadeh, P., M. Maddah, and T. Mirzaei, Using AI chatbots (eg, CHATGPT) in seeking health-related information online: the case of a common ailment. *Computers in Human Behavior: Artificial Humans*, 2025: p. 100127.
2. Ayre, J., E. Cvejic, and K.J. McCaffery, Use of ChatGPT to obtain health information in Australia, 2024: insights from a nationally representative survey. *Medical Journal of Australia*, 2025. 222(4): p. 210-212.
3. Choudhury, A. and H. Shamszare, Investigating the impact of user trust on the adoption and use of ChatGPT: survey analysis. *Journal of Medical Internet Research*, 2023. 25: p. e47184.
4. Ayo-Ajibola, O., et al., Characterizing the Adoption and Experiences of Users of Artificial Intelligence–Generated Health Information in the United States: Cross-Sectional Questionnaire Study. *Journal of Medical Internet Research*, 2024. 26: p. e55138.

[4] https://www.anthropic.com/claude/sonnet
[5] https://app.grammarly.com/

5. Kadian, A., et al., I Can't Take This Anymore! Understanding the Relationship Between Personality Traits and Tolerance of Generative AI Hallucinations. *IEEE Transactions on Engineering Management*, 2025.
6. Bates, D.W., et al., The potential of artificial intelligence to improve patient safety: a scoping review. *NPJ digital medicine*, 2021. 4(1): p. 54.
7. Davenport, T. and R. Kalakota, The potential for artificial intelligence in healthcare. *Future healthcare journal*, 2019. 6(2): p. 94-98.
8. Topol, E., Deep medicine: how artificial intelligence can make healthcare human again. 2019: Hachette UK.
9. Carayon, P. and P. Hoonakker, Human factors and usability for health information technology: old and new challenges. *Yearbook of medical informatics*, 2019. 28(01): p. 071-077.
10. Fraser, H., E. Coiera, and D. Wong, Safety of patient-facing digital symptom checkers. The Lancet, 2018. 392(10161): p. 2263-2264.
11. Ji, Z., et al., Survey of hallucination in natural language generation. *ACM computing surveys*, 2023. 55(12): p. 1-38.
12. Amann, J., et al., Explainability for artificial intelligence in healthcare: a multidisciplinary perspective. *BMC medical informatics and decision making*, 2020. 20(1): p. 310.
13. Longoni, C., A. Bonezzi, and C.K. Morewedge, Resistance to medical artificial intelligence. *Journal of consumer research*, 2019. 46(4): p. 629-650.
14. Bornstein, R.F., The dependent personality: developmental, social, and clinical perspectives. *Psychological bulletin*, 1992. 112(1): p. 3.
15. Giordano, V., et al., The impact of ChatGPT on human skills: A quantitative study on twitter data. *Technological Forecasting and Social Change*, 2024. 203: p. 123389.
16. Chu, Y. and P. Liu, Automation complacency on the road. *Ergonomics*, 2023. 66(11): p. 1730-1749.
17. Goh, A.Y., A. Hartanto, and N.M. Majeed, Generative artificial intelligence dependency: Scale development, validation, and its motivational, behavioral, and psychological correlates. *Computers in Human Behavior Reports*, 2025: p. 100845.
18. Lee, J.D. and K.A. See, Trust in automation: Designing for appropriate reliance. *Human factors*, 2004. 46(1): p. 50-80.
19. Turel, O., & Cui, T. (2026). Apologizing artificial intelligence: designing and evaluating effective AI apologies after errors. AI & SOCIETY, 1-18.
20. Kahneman, D., Attention and effort. Vol. 1063. 1973: Citeseer.
21. Lang, A., The limited capacity model of mediated message processing. Journal of communication, 2000. 50(1): p. 46-70.
22. Parasuraman, R. and V. Riley, Humans and automation: Use, misuse, disuse, abuse. *Human factors*, 1997. 39(2): p. 230-253.
23. Metzger, M.J. and A.J. Flanagin, Credibility and trust of information in online environments: The use of cognitive heuristics. Journal of pragmatics, 2013. 59: p. 210-220.
24. Turel, O. and S. Kalhan, Prejudiced against the machine? Implicit associations and the transience of algorithm aversion. MIS Quarterly, 2023. 47(4): p. 1369-1394.
25. Buçinca, Z., M.B. Malaya, and K.Z. Gajos, To trust or to think: cognitive forcing functions can reduce overreliance on AI in AI-assisted decision-making. *Proceedings of the ACM on Human-computer Interaction*, 2021. 5(CSCW1): p. 1-21.

26. Alarcon, G.M. and A. Capiola, Explicating the trust process for effective human interaction with artificial intelligence and machine learning systems. Frontiers in Computer Science, 2025. 7: p. 1662185.
27. Bickmore, T.W., et al., Patient and consumer safety risks when using conversational assistants for medical information: an observational study of Siri, Alexa, and Google Assistant. *Journal of medical Internet research*, 2018. 20(9): p. e11510.
28. Ye, J.-H., et al., The relationship between inert thinking and ChatGPT dependence: An I-PACE model perspective. *Education and information technologies*, 2025. 30(3): p. 3885-3909.
29. Abdelwanis, M., et al., Exploring the risks of automation bias in healthcare artificial intelligence applications: A Bowtie analysis. Journal of Safety Science and Resilience, 2024. 5(4): p. 460-469.
30. Walker HL, Ghani S, Kuemmerli C, Nebiker CA, Muller BP, Raptis DA, Staubli SM. Reliability of Medical Information Provided by ChatGPT: Assessment Against Clinical Guidelines and Patient Information Quality Instrument. *J Med Internet Res*. 2023;25:e47479. doi:10.2196/47479.
31. Yau JYS, Langdorf MI. Accuracy of Prospective Assessments of 4 Large Language Model Chatbot Responses to Patient Questions About Emergency Care: Experimental Comparative Study. *J Med Internet Res*. 2024;26:e60291. doi:10.2196/60291.
32. Kauttonen J, Rousi R, Alamaki A. Trust and Acceptance Challenges in the Adoption of AI Applications in Health Care: Quantitative Survey Analysis. *J Med Internet Res*. 2025;27:e65567. doi:10.2196/65567.
33. Wichmann J, Gesk TS, Leyer M. Acceptance of AI in Health Care for Short- and Long-Term Treatments: Pilot Development Study of an Integrated Theoretical Model. *JMIR Formative Research*. 2024;8:e48600. doi:10.2196/48600.
34. Sillence E, Briggs P, Harris PR, Fishwick L. How do patients evaluate and make use of online health information? *Soc Sci Med*. 2007;64(9):1853-1862. doi:10.1016/j.socscimed.2007.01.012.
35. Gaube S, Suresh H, Raue M, et al. Do as AI say: susceptibility in deployment of clinical decision-aids. *npj Digit Med*. 2021;4:31. doi:10.1038/s41746-021-00385-9.
36. Passi S, Vorvoreanu M. Overreliance on AI: Literature Review. Microsoft Research Technical Report MSR-TR-2022-12. 2022.
37. Steinfeld N. How Do Users Examine Online Messages to Determine If They Are Credible? An Eye-Tracking Study of Digital Literacy, Visual Attention to Metadata, and Success in Misinformation Identification. *Social Media + Society*. 2023;9(3). doi:10.1177/20563051231196871.
38. Ayers JW, Poliak A, Dredze M, et al. Comparing Physician and Artificial Intelligence Chatbot Responses to Patient Questions Posted to a Public Social Media Forum. *JAMA Intern Med*. 2023;183(6):589-596. doi:10.1001/jamainternmed.2023.1838.
39. Chi EH, Gumbrecht M, Hong L. Visual Foraging of Highlighted Text: An Eye-Tracking Study. In: Jacko J, ed. *Human-Computer Interaction, Part III*. HCII 2007. LNCS 4552. Springer. 2007;589-598. doi:10.1007/978-3-540-73110-8_64.
40. Doi T. Effective Highlighting Modes of Graphical User Interfaces in Visual Information Search. *Universal Access in the Information Society*. 2023;22:111-119. doi:10.1007/s10209-021-00827-x.